\documentclass[a4paper,11pt]{article}
\hbox{\tt PNUTP-23/A04} 
\usepackage{jcappub}
\usepackage[utf8]{inputenc}
\usepackage{graphicx,color,amssymb,amsmath,mathtools,cases,bm,float,tocbibind,subfigure,upgreek,longtable,geometry,blindtext,hyperref}
\usepackage[format=hang,font=small,textfont=it]{caption}
\allowdisplaybreaks[2]

\usepackage{pifont}  

\title{Numerical Investigation of Two-Dimensional Fokker-Planck Equation in Inflationary Models: Importance of Slow-Roll Parameters}
\author[a,b]{Deog Ki Hong,}
\author[a,b,1]{Jie Jiang,\note{Corresponding author}}
\author[c,d,e]{Dong-han Yeom}

\affiliation[a]{Center for Cosmological Constant Problem, Pusan National University, \\Busan 46241, Republic of Korea}
\affiliation[b]{Department of Physics, Pusan National University, \\Busan 46241, Republic of Korea}
\affiliation[c]{Department of Physics Education, Pusan National University, \\Busan 46241, Republic of Korea}
\affiliation[d]{Research Center for Dielectric and Advanced Matter Physics, Pusan National University, \\Busan 46241, Republic of Korea}
\affiliation[e]{Leung Center for Cosmology and Particle Astrophysics,
National Taiwan University, \\Taipei 10617, Taiwan}
\date{}
\emailAdd{dkhong@pusan.ac.kr}
\emailAdd{jiejiang@pusan.ac.kr}
\emailAdd{innocent.yeom@gmail.com}

\abstract{
In this study, we generalize the Fokker-Planck equation to two-dimensional cases, including potential functions with periodic boundary conditions and piecewise-defined structures, to analyze the probability distribution in multi-field inflationary models. We employ the spectral method for spatial derivatives and the Crank-Nicolson method for the time evolution to solve the equation numerically for the slow-roll inflation. We find that the distribution in the Fokker-Planck equation was determined by the two-dimensional potential combined slow-roll parameters. And the volume weighting effect makes the distribution in the Fokker-Planck Volume equation is determined by the potential.
}

\begin{document}


\maketitle


\section{Introduction}

In recent years, the study of cosmological inflation has garnered significant attention in the field of theoretical physics. Inflation is a scenario of an rapid expansion of space during the very early universe, right after Big Bang, that explains naturally the various observed features of our universe, such as the uniformity of the cosmic microwave background (CMB)~\cite{Guth:1980zm, Linde:1981mu, Albrecht:1982wi}, and furthermore it provides the generation of primordial density fluctuations, almost gaussian and scale-invariant, that lead to the formation of large-scale structures observed in the universe~\cite{Mukhanov:1981xt}. Inflationary models are based on scalar fields, known as inflatons, which drive the accelerated expansion of the universe. The evolution of these scalar fields is governed by their potential energy, and the slow-roll approximation is often used to simplify the analysis of these models.

The stochastic approach to inflation represents a significant theoretical development, as it provides a framework for investigating the quantum fluctuations that arise inevitably  during the evolution of the early universe. By introducing a probabilistic element to the dynamics of inflation, the stochastic approach reveals the inherent uncertainty in the evolution of the universe and highlights the need for a more nuanced understanding of the origins of cosmic structure. Stochastic inflation serves as a powerful tool for reconciling the predictions of inflation theory with the principles of quantum mechanics, as it illuminates the role that quantum fluctuations play in the generation of density perturbations and the resulting anisotropies in the cosmic microwave background. The incorporation of stochastic effects in inflationary models has led to a refinement of the theoretical framework, providing a more complete picture of the dynamics of the early universe and the emergence of structure therein.

One important aspect of studying inflation is the probability distribution of scalar fields in different regions of the inflaton potential. In other words, one may ask about the probability distribution of the \textit{multiverse}. We assume that the wavelength of fluctuations is sufficiently larger than the Hubble radius, and hence, one can approximate that each Hubble patch is a homogeneous and isotropic universe. In this background, the Fokker-Planck equation is a widely used tool for describing the evolution of probability distributions in such systems, incorporating the stochastic effects that arise from quantum fluctuations \cite{Starobinsky:1986fx,Vennin:2015hra}. However, it is crucial to ensure that the slow-roll conditions are satisfied for the Fokker-Planck equation to be applicable. In this paper, we present a numerical investigation of the evolution of probability distributions in two cases: one with the slow-roll condition slightly violated and the other with the slow-roll condition marginally satisfied.

The Fokker-Planck equation has recently been examined in relation to the cosmological constant problem, leading to the proposal of the intriguing Self-Organized Localization (SOL) scenario \cite{Giudice:2021viw,Jung:2021cps}. This scenario suggests that our universe likely resides at a critical point, where observed parameters are not constants but functions determined by assisted fields. However, the justification of the SOL scenario hinges on the assignment of probabilities beyond the critical point, posing challenges to the use of the Fokker-Planck equation. Exploring alternative avenues, genuine quantum cosmology based on the Wheeler-DeWitt equation \cite{DeWitt:1967yk} offers prospects for addressing this issue and merits further investigation.

To achieve accurate and stable results, we employ the spectral method \cite{trefethen2000spectral} in the spatial direction and the Crank-Nicolson method \cite{crank_nicolson_1947} in the temporal direction for our simulations. We also explore the influence of volume weighting on the probability distribution and the role of slow-roll parameters on the inflaton potential. Our study sheds light on the behavior of probability distributions in different scenarios and contributes to the understanding of the dynamics of such systems.

The paper is structured as follows. In Section 2, we provide an introduction to the two-field Fokker-Planck equation, establishing the basis for our analysis. Section 3 outlines our numerical methods, including the spectral method for spatial derivatives and the Crank-Nicolson method for temporal evolution. We also discuss the selection of model parameters and initial conditions. Our results for the two investigated cases are presented. Firstly, we examine the smooth two-field case, where the slow-roll condition is met, and analyze the profiles of $c_1$, the coefficient of the leading term of the  Fokker-Planck equation in the slow-roll approximation, and the corresponding distribution $P$. Secondly, we investigate the non-smooth two-field case, observing that the potential's discontinuity results in a corresponding discontinuity in $c_1$, leading to discontinuous distributions $P$. In Section 4, we discuss the implications of our findings for the Self-Organized Localization scenario. Finally, Section 5 summarizes our results and discusses their significance in understanding inflationary models. We emphasize the importance of stochastic inflationary models in advancing our comprehension of the universe's origins, as they offer a robust framework for exploring the quantum-mechanical fluctuations that underpin the early universe's evolution. Additionally, they provide a means to reconcile inflation theory with the principles of quantum mechanics.

Throughout our investigation, we aim to provide a comprehensive understanding of the evolution of probability distributions in different inflationary scenarios. Our findings can be used for the studies on the dynamics of scalar fields in the early universe and their implications for the formation of large-scale structures and the cosmic microwave background.

\section{Two-field Fokker-Planck equation}

\subsection{Fokker-Planck equation without volume-weighting}

From the perspective of stochastic inflation, we can divide the scalar field into two parts. One is the short wavelength perturbations, the other is the long wavelength perturbations. During inflation, perturbations gradually exit the horizon and contribute to the long wavelength part. From the sub-horizon point of view, long wavelength variation can be seen as noise \cite{Fujita:2013cna}. So the equations of motions of scalar fields $\phi$ and $\chi$ with its potential $V(\phi, \chi)$ in the sub-horizon, with $a$ being the scale factor and $H$ the Hubble parameter, 

\begin{gather}
	\ddot{ \phi } + 3 H \dot{ \phi } - \frac{ \nabla ^ 2 }{ a ^ 2 } \phi + V _ { , \phi } = 0 ~ , \\
	\ddot{ \chi } + 3 H \dot{ \chi } - \frac{ \nabla ^ 2 }{ a ^ 2 } \chi + V _ { , \chi } = 0 ~ ,
\end{gather}

\noindent can be expressed as 

\begin{gather}
	\dot{ \phi } = - \frac{ V _ { , \phi } }{ 3 H } + \frac{ H ^ { 3 / 2 } }{ 2 \pi } \Gamma _ { \phi } ( t ) ~ , \\
	\dot{ \chi } = - \frac{ V _ { , \chi } }{ 3 H } + \frac{ H ^ { 3 / 2 } }{ 2 \pi } \Gamma _ { \chi } ( t ) ~ ,
\end{gather}

\noindent where we use the homogeneity approximation and slow-roll conditions.  The noise terms satify

\begin{equation}
	\left\langle \Gamma _ { \phi } ( t ) \Gamma _ { \phi } ( t ^ { \prime } ) \right\rangle = \delta ( t - t ^ { \prime } ) ~ , \qquad \left\langle \Gamma _ { \chi } ( t ) \Gamma _ { \chi } ( t ^ { \prime } ) \right\rangle = \delta ( t - t ^ { \prime } ) ~ , \qquad \left\langle \Gamma _ { \phi } ( t ) \Gamma _ { \chi } ( t ^ { \prime } ) \right\rangle = 0 ~ .
\end{equation}

The noise of the random walk comes from the exited long wavelength mode perturbation. So, no matter whether the two fields couple or not, their noise only has the contribution from themselves.

In this study, we employ cosmic time as the temporal parameter. It has been noted in literature that the utilization of the $e$-folding number $N$ as the temporal variable is advisable, as it obviates the need for corrections to the noise terms \cite{Pattison:2019hef}. However, we demonstrate in our investigation that these two time variables are interchangeable, and we adopt cosmic time in our analysis.

After lengthy calculations, we can get the Fokker-Planck equation for two-fields
for the normalized probability distribution $P_{\rm PF}(\phi,\chi,t)$, given as 
 
\begin{equation}\label{FPE}
	\frac{ \partial P _ { \rm FP } }{ \partial t } = \frac { \partial } { \partial \phi } \left[ \frac { V _ { , \phi } } { 3 H } P _ { \rm FP } + \frac { H ^ { 3 ( 1 - \beta ) } } { 8 \pi ^ 2 } \frac{ \partial }{ \partial \phi } \left( H ^ { 3 \beta } P _ { \rm FP } \right) \right] + \frac { \partial } { \partial \chi } \left[ \frac { V _ { , \chi } } { 3 H } P _ { \rm FP } + \frac { H ^ { 3 ( 1 - \beta ) } } { 8 \pi ^ 2 } \frac{ \partial }{ \partial \chi } \left( H ^ { 3 \beta } P _ { \rm FP } \right) \right] ~ ,
\end{equation}

\noindent where $ \beta = 1 / 2 $ corresponds to the Stratonovich version of stochastic analysis, $ \beta = 1 $ corresponds to the Ito interpretation. It has been demonstrated that the stochastic differential equation governing inflationary dynamics should be discretized in accordance with Ito's approach to maintain one-loop consistency with quantum field theory \cite{Tokuda:2017fdh, Pinol:2020cdp}. In this study, we adopt the Stratonovich formulation; however, we will also examine the Ito formulation for comparative analysis towards the end, aiming to explore the disparities stemming from discretization methodologies \cite{Assadullahi:2016gkk}.

\noindent For convenience, we introduce a set of slow-roll parameters for the potential as

\begin{gather}
	\epsilon _ { \phi \phi } = \frac{ M _ { \rm Pl } ^ 2 }{ 2 } \left( \frac{ V _ { , \phi } }{ V } \right) ^ 2 ~ , \\
	\epsilon _ { \chi \chi } = \frac{ M _ { \rm Pl } ^ 2 }{ 2 } \left( \frac{ V _ { , \chi } }{ V } \right) ^ 2 ~ , \\
	\eta _ { \phi \phi } = M _ { \rm Pl } ^ 2 \frac{ V _ { , \phi \phi } }{ V } ~ , \\
	\eta _ { \chi \chi } = M _ { \rm Pl } ^ 2 \frac{ V _ { , \chi \chi } }{ V } ~ ,
\end{gather}

\noindent and express Eq. \eqref{FPE} with the slow-roll parameters

\begin{equation}\label{PFP_in_slowroll}
	P _ { {\rm FP}, t } = c _ { 1 _ {\rm FP} } P _ { \rm FP }+ c _ { 2 } P _ { {\rm FP}, \phi } + c _ { 3 } P _ { {\rm FP}, \chi } + c _ { 4 } ( P _ { {\rm FP}, \phi \phi } + P _ { {\rm FP}, \chi \chi } ) ~ ,
\end{equation}

\noindent where

\begin{gather}
	c _ { 1 _ {\rm FP } } = \frac{ V ^ { 1 / 2 } }{ \sqrt{ 3 } M _ { \rm Pl } } \left( \eta _ { \phi \phi } + \eta _ { \chi \chi } - \epsilon _ { \phi \phi } - \epsilon _ { \chi \chi } \right) + \frac{ V ^ { 3 / 2 } }{ 32 \sqrt{ 3 } \pi ^ 2 M _ { \rm Pl } ^ 5 } \left( \eta _ { \phi \phi } + \eta _ { \chi \chi } + \epsilon _ { \phi \phi } + \epsilon _ { \chi \chi } \right) ~ , \\
	c _ { 2 } = \mathrm{sign}(V _ \phi) \frac{ \sqrt{ 6 } V ^ { 1 / 2 } }{ 3 } \sqrt{ \epsilon _ { \phi \phi } } + \mathrm{sign}(V _ \phi) \frac{ \sqrt{ 6 } V ^ { 3 / 2 } }{ 32 \pi ^ 2 M _ { \rm Pl } ^ 4 } \sqrt{ \epsilon _ { \phi \phi } } ~ , \\
	c _ { 3 } = \mathrm{sign}(V_\phi) \frac{ \sqrt{ 6 } V ^ { 1 / 2 } }{ 3 } \sqrt{ \epsilon _ { \chi \chi } } + \mathrm{sign}(V_\phi) \frac{ \sqrt{ 6 } V ^ { 3 / 2 } }{ 32 \pi ^ 2 M _ { \rm Pl } ^ 4 } \sqrt{ \epsilon _ { \chi \chi } } ~ , \\
	c _ { 4 } = \frac{ V ^ { 3 / 2 } }{ 24 \sqrt{ 3 } \pi ^ 2 M _ { \rm Pl } ^ 3 } ~ .
\end{gather}

\subsection{Fokker-Planck equation with volume-weighting}

In our investigation, we evaluate rather the volume-weighted probability distribution $P_{\rm VW}$ instead of the normalized distribution $P_{\rm FP}$.  
The volume weighted probability distribution entails the meticulous evaluation of the volume-weighted dispersion of field values, obtained by computing the ensemble average of $ \langle e ^ { 3 H t } \rangle $ over a collection of random walks. To denote this dispersion, we shall denote it as $ P _ {\rm VW}$, wherein it exhibits a proportionality to $ e ^ { 3 H t } P _ { \rm FP } $. In order to proceed, we shall substitute $ P _ { \rm FP } = \alpha e ^ { - 3 H t } P _ {\rm VW}$ into Equation \eqref{FPE} to find 

\begin{align}\label{VWE}
	\frac{ \partial P _ {\rm VW}}{ \partial t } = & \frac { \partial } { \partial \phi } \left[ \frac { V _ { , \phi } } { 3 H } P _ {\rm VW} + \frac { H ^ { 3 / 2 } } { 8 \pi ^ 2 } \frac{ \partial }{ \partial \phi } \left( H ^ { 3 / 2 } P _ {\rm VW} \right) \right] + \frac { \partial } { \partial \chi } \left[ \frac { V _ { , \chi } } { 3 H } P _ {\rm VW} + \frac { H ^ { 3 / 2 } } { 8 \pi ^ 2 } \frac{ \partial }{ \partial \chi } \left( H ^ { 3 / 2 } P _ {\rm VW} \right) \right] \nonumber \\
	& + 3 H P _ {\rm VW} ~ ,
\end{align}

\noindent where the last term indicates the volume-weighted effect~\cite{Nakao:1988yi,Nambu:1989uf,Linde:1993xx,Nambu:1988je}. We can also express Eq. \eqref{VWE} with the slow-roll parameters,

\begin{equation}\label{PVW_in_slowroll}
	P _ { {\rm VW}, t } = c _ { 1 _ {\rm VW} } P _ { \rm VW }+ c _ { 2 } P _ { {\rm VW}, \phi } + c _ { 3 } P _ { {\rm VW}, \chi } + c _ { 4 } ( P _ { {\rm VW}, \phi \phi } + P _ { {\rm VW}, \chi \chi } ) ~ .
\end{equation}

\noindent In this case, the coefficient of $ P _ {\rm VW} $ in Eq. \eqref{PVW_in_slowroll} is

\begin{equation}\label{c1_vw}
	 c _ { 1 _ {\rm VW} } = \frac{ V ^ { 1 / 2 } }{ \sqrt{ 3 } M _ { \rm Pl } } \left( 3 + \eta _ { \phi \phi } + \eta _ { \chi \chi } - \epsilon _ { \phi \phi } - \epsilon _ { \chi \chi } \right) + \frac{ V ^ { 3 / 2 } }{ 32 \sqrt{ 3 } \pi ^ 2 M _ { \rm Pl } ^ 5 } \left( \eta _ { \phi \phi } + \eta _ { \chi \chi } + \epsilon _ { \phi \phi } + \epsilon _ { \chi \chi } \right) \, .
\end{equation}

\section{Numerical evaluation}

For our simulation of the evolution of probability, we employ the spectral method in the spatial direction and the Crank-Nicolson method in the temporal direction. The spectral method offers an advantage over traditional differentiation methods in computing spatial derivatives. While differentiation methods rely on the local information from adjacent points to infer the derivative at a specific point, the spectral method utilizes the global information from the entire space, resulting in greater accuracy. Furthermore, the Crank-Nicolson method is consistently stable for large-time steps. 

We employ periodic boundary conditions to avoid potential difficulties in calculations caused by other boundary conditions such as zero boundary conditions, which may result in issues like half-wave losses. We can do the integration of Eq. \eqref{FPE} over the field space, and get

\begin{align}
	& \int \frac{ \partial P _ {\rm FP} }{ \partial t } d\phi d\chi \nonumber \\
 = &\int \left. \left[ \frac { V _ { , \phi } } { 3 H } P _ {\rm FP} + \frac { H ^ { 3 / 2 } } { 8 \pi ^ 2 } \frac{ \partial }{ \partial \phi } \left( H ^ { 3 / 2 } P _ {\rm FP} \right) \right] \right| ^ { \phi _ { \rm max } } _ { \phi _ { \rm min } } d\chi + \int \left.  \left[ \frac { V _ { , \chi } } { 3 H } P _ {\rm FP} + \frac { H ^ { 3 / 2 } } { 8 \pi ^ 2 } \frac{ \partial }{ \partial \chi } \left( H ^ { 3 / 2 } P _ {\rm FP} \right) \right] \right| ^ { \chi _ { \rm max } } _ { \chi _ { \rm min } } d\phi \nonumber \\
= & ~ 0 \label{anti}\, .
\end{align}

\noindent By the periodic boundary conditions, the antiderivative of the integral of the left-hand side of Eq.~(\ref{anti}) vanishes automatically. As a result, the integration of the probability, denoted by $\int P _ {\rm FP} d\phi d\chi$, remains conserved. This can serve as a criterion for verifying the accuracy of calculations.

\subsection{Two-field model}

In this section, we analyze a two-fields inflation model described by the potential function $V(\phi, \chi)$ given as 

\begin{equation}
	V(\phi, \chi) = 3H_0M_{\rm{Pl}}^2 + gf^4\left(\cos\frac{\phi}{f}\cos\frac{2\phi}{f} + \cos\frac{\chi}{f}\cos\frac{2\chi}{f} + \cos\frac{\phi}{f}\cos\frac{\chi}{f}\right).
\end{equation}

\subsubsection{Classical region}

\begin{figure}[hbt]
     \centering
     \includegraphics[width=\textwidth]{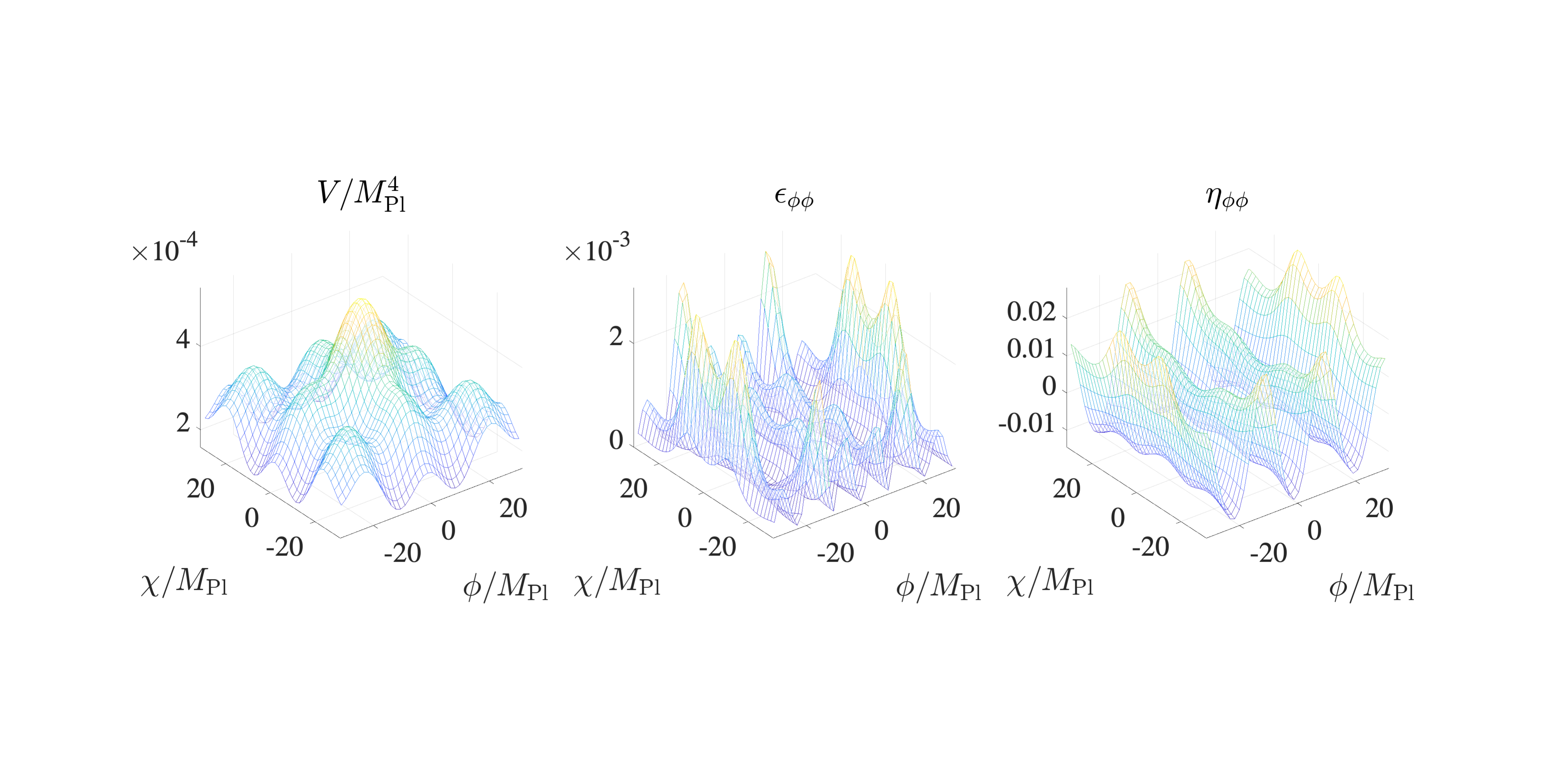}
     \caption{Parameters of a two-field model.}
     \label{pb_double_potential}
\end{figure}

We choose the parameters $H_0 = 1\times10^{-2}M_{\rm{Pl}}$, $f = 10M_{\rm{Pl}}$, and $g = 8\times10^{-9}$. For our choice of parameters, which satisfy the CbQ (Classical-beat-Quantum) and slow-roll conditions, $ H ^ 3 < | V ^ { \prime } | < M_{\rm Pl} H ^ 2 $, which means that the evolution of scalar fields is in a classical (C) region. The inflaton fields $\phi$ and $\chi$ are confined to the range $-\pi < \phi/f < \pi$ and $-\pi < \chi/f < \pi$ to enforce periodic boundary conditions. We assume an initial distribution of field values that follows a constant distribution, implying an equal likelihood for each field value. The potential $V$ and the slow-roll parameters are depicted in Fig. \ref{pb_double_potential}. As the two fields are symmetric in this model, only one set of slow-roll parameters corresponding to one field is plotted. The figure shows that the slow-roll conditions are satisfied, validating the assumptions made in the Fokker-Planck equation.

\begin{figure}[hbt]
     \centering
     \includegraphics[width=0.6\textwidth]{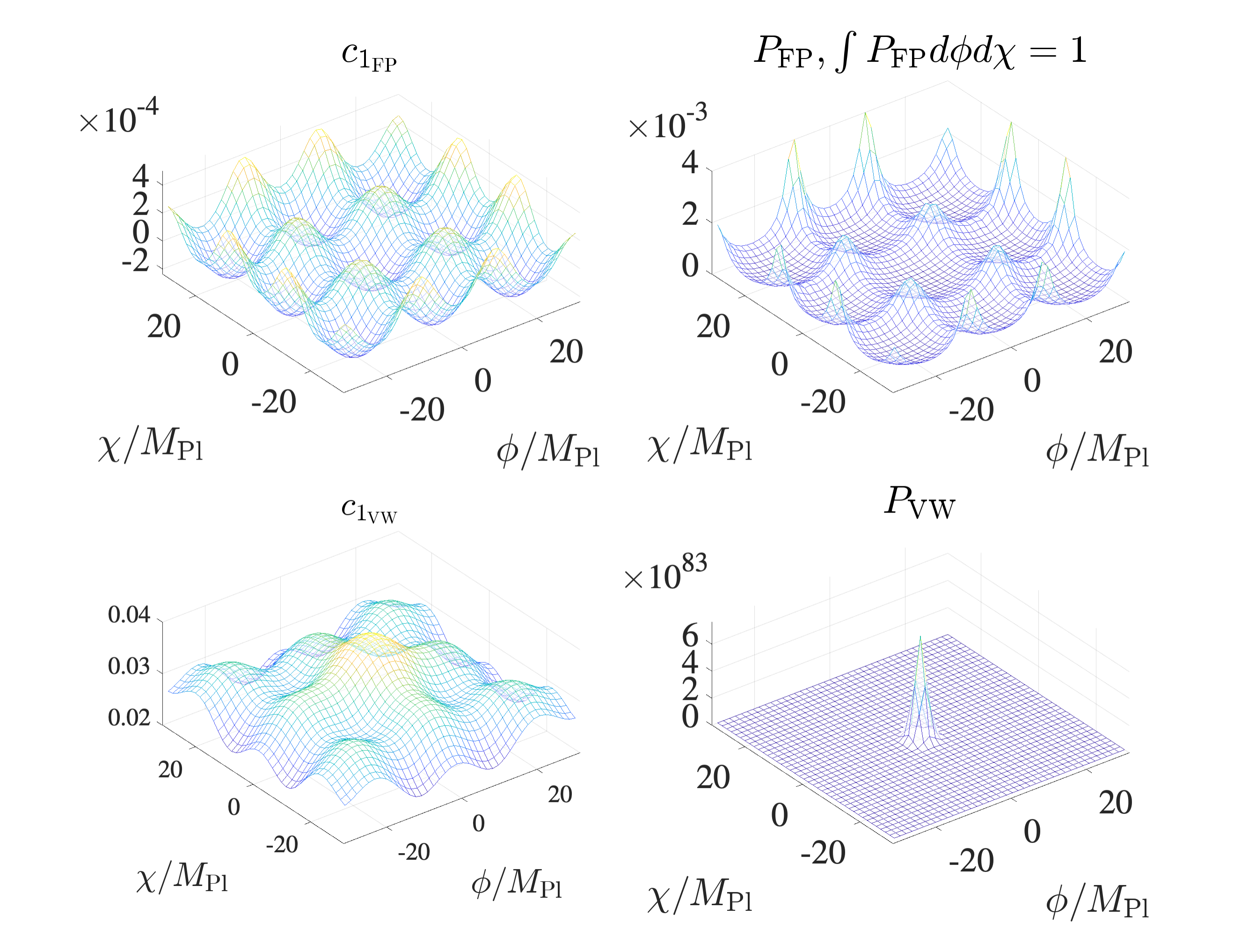}
     \caption{Probability distribution $ P _ { \rm FP } $ and volume distribution $ P _ { \rm VW } $ and corresponding $c_1$ parameter in two-fields case.($C$ regime)}
     \label{pb_double_c}
\end{figure}

Next, we explore the relationship between the probability distribution $P_{\rm{FP}}$ and the volume distribution $P_{\rm{VW}}$, along with their respective $c_1$ parameters in Figure \ref{pb_double_c}. It is evident that the integral $\int P_{\rm{FP}} d\phi d\chi$ remains conserved in the presence of periodic fields, affirming the accuracy of our numerical algorithm.

The upper panel of Figure \ref{pb_double_c} illustrates the correlation between the profile of the probability distribution, denoted as $P_{\rm{FP}}$, and $c_{1_{\rm {FP}}}$ in the C region. Without the influence of volume weighting, the probability distribution $P_{\rm{FP}}$ coincides with the local maximum of $c_{1_{\rm {FP}}}$, indicating the local minimum of the potential. Moreover, the global maximum of $P_{\rm{FP}}$ at the figure's edge corresponds to the potential's global minimum, as depicted in Figure \ref{pb_double_potential}. This suggests that in the classical region, absent the effect of volume weighting, the probability distribution $P_{\rm{FP}}$ tends to be drawn towards the potential's global minimum.

A similar correlation is observed in the lower panel of Figure \ref{pb_double_c}, which illustrates the relationship between the volume distribution $P_{\rm{VW}}$ and $c_{1_{\rm{VW}}}$ in the C region. Here, the global maximum of $P_{\rm{VW}}$ coincides with the global maximum of $c_{1_{\rm{VW}}}$, which also represents the maximum of the potential. This is attributed to the dominance of the classical effect. Due to the classical effect domination, the volume weighting effect causes the distribution $P_{\rm{VW}}$ at the global maximum of the potential to undergo the most significant exponential expansion, effectively dominating over the entire field space. Consequently, a single prominent peak emerges in the distribution, as other field values are substantially smaller in comparison.

To elucidate this occurrence, we consider the equation. At the point of the local maximum, the derivative becomes zero, leading to the equation $P_{,t} = c_1P + c_4(P_{,\phi\phi} + P_{,\chi\chi}) \approx c_1P$, where the last approximation holds because $|c_4|$ is smaller than $|c_{1_{\rm{VW}}}| $. This implies that the local maximum of $c_1$ governs the location of the local maximum of the distribution.

\subsubsection{Quantum region}

In this subsection, we plot with $H_0 = 1\times10^{-2}M_{\rm{Pl}}$, $f = 10M_{\rm{Pl}}$, and $g = 4\times10^{-11}$. Where in this case, $ H ^ 4 / M_{\rm Pl} < | V ^ { \prime } | <  H ^ 3 $, which means this is in $ \text{Quantum} + \text{Volume} $ ($\text{QV}$) region. 

\begin{figure}[hbt]
     \centering
     \includegraphics[width=0.6\textwidth]{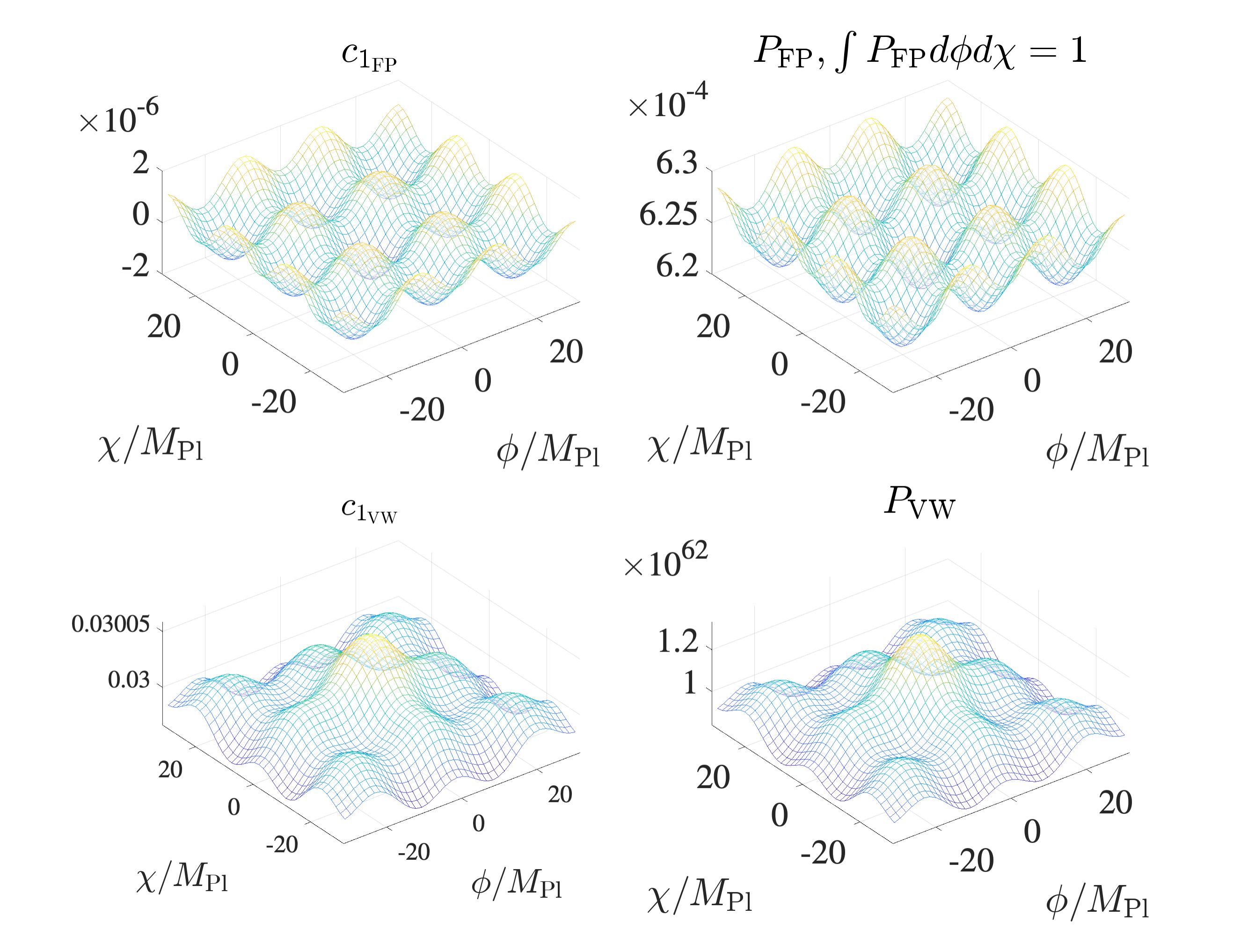}
     \caption{Probability distribution $ P _ { \rm FP } $ and volume distribution $ P _ { \rm VW } $ and corresponding $c_1$ parameter in two-fields case.($Q V$ regime)}
     \label{pb_double-QV}
\end{figure}

Figure \ref{pb_double-QV} exhibits an unexpected disparity between the profile of the probability distribution $P_{\rm{FP}}$ and the potential $V$. The probability does not have its maximum at the minimum of the potential. However, we observe that the probability distribution $P_{\rm{FP}}$ has the same profile with that of $c_{1_{\rm {FP}}}$. A similar correlation is observed between the volume distribution $P_{\rm{VW}}$ and $c_{1_{\rm{VW}}}$. Hence, the location of the local maximum of the distribution is determined by $c_1$. And we also see the same result in $ \text{Quantum} ^ 2 + \text{Volume} $ ($\text{Q}^2 \text{V}$), where $ H ^ 4 / M_{\rm Pl} > | V ^ { \prime } | $.

One aspect we wish to underscore pertains to the Quantum realm, where classical effect dominance is obviated. Consequently, the volume distribution $P_{\rm VW}$ no longer exhibits a conspicuous peak corresponding to the global maximum of the potential. Instead, it mirrors the configuration of $c_{1_{\rm{VW}}}$, representing the potential contribution in the context of volume weighting.

\subsection{Two-field non-smooth model}

To investigate the discontinuity in the potential and its influence on the distribution in \cite{Giudice:2021viw}, we analyze a two-fields non-smooth inflation model described by the potential function $V(\phi, \chi)$ given by:

\begin{equation}
	V(\phi, \chi) = 3H_0M_{\rm{Pl}}^2 + g f^4\left[ a \left( \left| \frac{\phi }{ f } - \left\lfloor \frac{\phi }{ f } + \frac{1 }{ 2 } \right\rfloor \right| + \left| \frac{\chi }{ f } - \left\lfloor \frac{\chi }{ f } + \frac{1 }{ 2 } \right\rfloor \right| \right) + b \left( | \phi | + | \chi | \right)\right],
\end{equation}
where $ \lfloor x \rfloor $ is the floor function, $H_0 = 1\times10^{-2}M_{\rm{Pl}}$, $f = 10M_{\rm{Pl}}$, $g = 1 \times 10 ^ {-5} $, $ a= 20 $ and $ b = 1$. The inflaton fields $\phi$ and $\chi$ are confined to the range $-2 < \phi/f < 2$ and $-2 < \chi/f < 2$ to enforce periodic boundary conditions.

\begin{figure}[hbt]
     \centering
     \includegraphics[width=\textwidth]{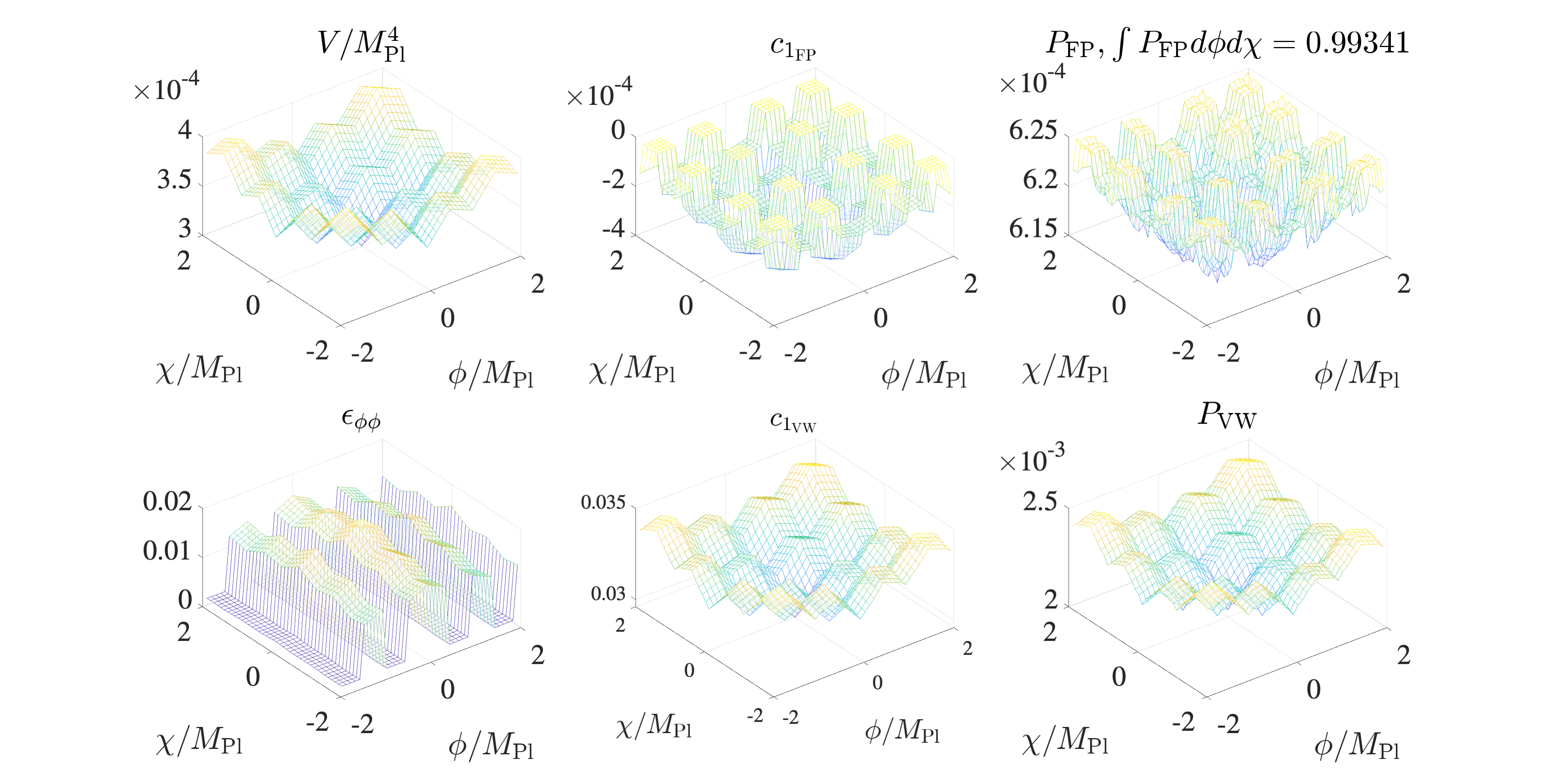}
     \caption{Potential, first slow-roll parameter, probability distribution $ P _ { \rm FP } $, volume distribution $ P _ { \rm VW } $ and corresponding $c_1$ parameter in two-fields non-smooth case.}
     \label{pb_piecewise_potential}
\end{figure}

In Fig. \ref{pb_piecewise_potential} we observe a similar result to what was shown in the previous section. The discontinuity in the potential leads to a discontinuity in $ c _ 1 $, which, in turn, determines the profile of the distributions.
This observation suggests that the nature of the potential plays a crucial role in shaping the distribution. The discontinuities in the potential generate corresponding discontinuities in the $ c _ 1 $ parameters, influencing the behavior of the distributions $ P $.

We illustrate the behavior of the Fokker-Planck equation with and without the volume weighting effect. In this context, the non-smooth nature of the potential and its associated non-smooth slow-roll parameter contribute to the distinctive blocky raised structure in $c_{1_{\rm {FP}}}$. Consequently, this non-smoothness propagates to the distribution $ P _ { \rm FP } $, resulting in a corresponding blocky raised shape. In the Fokker-Planck Volume equation, $c_{1_{\rm {VW}}}$ is primarily determined by the potential, leading to a similar non-smooth shape in the distribution $ P _ { \rm VW } $.


\subsection{Summary of numerical results}

Our analysis shows the slow-roll parameters determine the peak position of the probability distribution in both smooth  and non-smooth potentials of the inflationary models. 

In the first section, we examined a two-fields inflation model with a specific potential function. The model's potential and slow-roll parameters were analyzed to find that the satisfaction of slow-roll conditions is essential for the Fokker-Planck equation. The disparity between the profile of the probability distribution $P_{\rm{FP}}$ and the inverse of potential $V$ is particularly pronounced in this model. Furthermore, we investigated the relationship between probability distribution $P_{\rm{FP}}$ and volume distribution $P_{\rm{VW}}$ along with their corresponding $c_1$ parameters. The local maximum of $c_1$ was found to determine the location of the local maximum in the distribution $ P $.

In the second section, we delved into a two-fields non-smooth inflation model characterized by a piecewise-defined potential function. The potential exhibited discontinuities, leading to corresponding discontinuities in $c_1$, which affect the shape of the distributions. Figure \ref{pb_piecewise_potential} illustrates this phenomenon. Understanding the role of the potential discontinuities in determining the probability distribution is crucial for the inflation dynamics and its consequences for the early universe.

In both the classical and quantum cases, the fact that $P_{\rm{VW}}$ is peaked at the maximum of the potential is indeed relevant to SOL. The key difference lies in how the other local maximum of the potential shows in the distribution.

In the classical case, the volume weighting effect causes the distribution $P_{\rm{VW}}$ at the global maximum of the potential to experience the most significant exponential expansion, effectively dominating over the entire field space. This results in a single prominent peak in the distribution, as other field values are substantially smaller in comparison.

In the quantum case, where $P_{\rm{VW}}$ mirrors the shape of the potential, the probability distribution spreads out across the multiple local maxima of the potential. While the global maximum still retains importance, the existence of these other local maxima means that they are not entirely ignored in the distribution.

Our findings shed light on the intricate interplay between the features of the  potential, especially $ c_ 1$, the leading term in the Fokker-Planck equation, expanded in the slow-roll parameters,  and the resulting distribution during inflationary epochs. Such insights might enhance our understanding of inflationary cosmology and the evolution of the early universe.

We also plot the corresponding Ito version figures which we show in the paper, they have the similar result so we do not show them in the paper. This result agree with the previous paper that the stochastic way does not matter the final result too much.

\section{Applications for the Self-Organized Localization scenario}

Recently, the Fokker-Planck equation was investigated in the context of the cosmological constant problem. Although there is no definite consensus on this difficult problem, one interesting idea was proposed recently, which is called the \textit{Self-Organized Localization} (SOL) scenario \cite{Giudice:2021viw,Jung:2021cps}.

The bottom line of this idea is that some observed parameters (which look finely tuned) are indeed not constant parameters  but functions determined by assisted fields. The underlying potential structures can have a critical behavior, where they can show two different field regions that correspond to different phases. The assertion of the SOL scenario is that our universe probably finds observed parameters at the critical point. The question is, how can it be justified?

Here, we need some principles of quantum cosmology, i.e., the way to assign probabilities over the multiverse. In the original SOL paper, the authors introduced the volume-weighted Fokker-Planck equation for this purpose. Near the critical point, the potential structure might be too steep, and hence, it might be reasonable to restrict the field domain of the Fokker-Planck equation, such that the domain is close to the critical point. Then, in many cases, near the boundary of the integration domain, i.e., near the critical point, the potential energy is larger than the other regions; hence, if we consider the volume-weighted Fokker-Planck equation, a sharp peak of the probability function will appear near the critical point. This is the SOL paradigm that explains why our universe has started from such a critical point.

The SOL paradigm is an intriguing idea but still has some  caveats. We have assumed an underlying potential structure that has a phase transition. In principle, we should be able to assign the probability for the entire field space. In other words, unless there exists a fundamental reason why we should restrict parameters inside the integration domain, one can ask about the probability beyond the critical point. However, due to the steepness of the potential near the phase transition, it is very doubtful whether the slow-roll conditions, i.e., the fundamental assumption for the use of the Fokker-Planck equation, globally hold or not. Therefore, if we can not assign the probability for the global field space, the Fokker-Planck equation might not be justifiable and the SOL paradigm might be in question. 

The answer to this question is howver beyond the scope of this paper, but one may revisit the ideas of genuine quantum cosmology based on the Wheeler-DeWitt equation. The final result might depend on the use of the boundary conditions, e.g., using the no-boundary wave function or the tunneling wave function. However, if we use this method, we can go beyond the slow-roll approximation and ask about the validity of the SOL scenario. We leave this interesting topic for future investigations.

\section{Conclusion}

In this paper, we have analyzed a two-fields inflation model described by a potential function with periodic boundary conditions and a non-smooth inflation model with a piecewise-defined potential. The slow-roll conditions are chosen to satisfy in both models, as they are the essential assumptions in the Fokker-Planck equation. The probability distribution $P_{\rm{FP}}$ was found to exhibit unexpected disparities with the potential $V$, but a correlation between the local maximum of $P_{\rm{FP}}$ and the $c_{1_{\rm FP}}$ parameter was observed. In the volume-weighted case, it also shows same phenomenon, namely the profile of $ c_{1_{\rm VW}} $ determines the local maximum of $P_{\rm{VW}}$. Finally, we find that the non-smooth potential model shows discontinuities in both the potential and $c_1$ parameters, affecting the shape of the distributions.

Finally we have discussed the implications of our findings in the context of the Self-Organized Localization (SOL) scenario, which addresses the cosmological constant problem. The SOL scenario suggests that observed parameters are not constants but functions determined by assisted fields, and our universe likely has started from a critical point near a phase transition. However, using the Fokker-Planck equation to assign probabilities over the entire field space might be problematic due to the steepness of the potential near the critical point. The justification of the SOL scenario depends on whether slow-roll conditions hold globally or not, which is still an open question.

We must be careful to apply the Fokker-Planck equation to quantum cosmological situations, such as the self-organized localization scenario. We might need a better quantum cosmological tool to assign probabilities, even though the potential structure does not satisfy the slow-roll conditions. Perhaps, canonical quantization might provide such a tool, but one can ask whether that method will be useful to revive the ideas of the self-organized localization scenario. We leave this topic for future investigations.

To address these uncertainties, future investigations could explore genuine quantum cosmology based on the Wheeler-DeWitt equation and consider different boundary conditions, such as the no-boundary wave function or the tunneling wave function. This approach may allow us to go beyond the slow-roll approximation and provide a more comprehensive understanding of the validity of the SOL scenario and its implications for inflationary cosmology.

During our study, we have developped a strong numerical tool to solve the Fokker-Planck equation. One may use this to study interesting issues such as stochastic effects of inflationary scenarios which have multiple field directions. We also leave this topic for future investigations.

\acknowledgments
This work was supported by the National Research Foundation of Korea (NRF) grant funded by the Korea government (MSIT) (2021R1A4A5031460). 
The authors would like to thank Sunghoon Jung, Misao Sasaki and Zihan Zhou for the useful discussion. DY was further supported by the National Research Foundation of Korea (Grant No. : 2021R1C1C1008622).

\bibliographystyle{JHEP}
\bibliography{bibliography_FPE.bib}{} 


\end{document}